\documentclass{PoS}
\usepackage{lineno}
\usepackage{amsmath}
\title{Potential constrains on Lorentz invariance violation from the HAWC TeV gamma-rays}

\ShortTitle{Potential LIV constrains from HAWC}

\author{\speaker{H. Mart\'inez-Huerta}  for the HAWC 
Collaboration \thanks{For a complete author list, see 
http://www.hawc-observatory.org/collaboration/icrc2017.php}\\
        Departamento de F\'isica, Centro de Investigaci\'on y de Estudios Avanzados del I.P.N.,\\ 
        Apdo. Post. 14-740, 07000, Ciudad de M\'exico, M\'exico\\
        E-mail: \email{hmartinez@fis.cinvestav.mx}}

\abstract{
Astrophysical scenarios provide a unique opportunity to test the possible signatures of Lorentz Invariance Violation (LIV) due to the high energies and the very long distances they involve. An isotropic correction to the photon dispersion relation, by hypothetical Lorentz invariance violation, has a consequence that photons of sufficient energy are unstable and decay very fast. The High Altitude Water Cherenkov (HAWC) observatory is sensitive to gamma-rays in the 100 GeV to 100 TeV energy range, making it a very useful tool to study LIV. In this work we present potential stringent limits for the LIV energy scale at first and second order correction by the potential observations of primary very high energy photons in HAWC energy range.
}

\FullConference{35th International Cosmic Ray Conference $–$ ICRC217$-$\\
		10-20 July, 2017\\
		Bexco, Busan, Korea}

\begin{document}

\section{Introduction}

Lorentz symmetry stands as one of the cornerstones of fundamental physics. Nonetheless, as for any other fundamental principle, exploring its limits of validity has been an important motivation for theoretical and experimental research in the past.
In addition some  Lorentz Invariance Violation (LIV) can be motivated by beyond the Standard Model theories, such as quantum gravity and string theories
(see for instance Refs.~\cite{NAMBU,Bluhm,Pot,ALFARO,QG1,QG2,QG3,QG4,QG5,Gian,SME}).  Among the possibilities, some signatures of such violation could emerge from the corrected free particle dispersion relation and tested by astrophysical sources of very high energy  (VHE) photons 
\cite{GLASHOW,GLASHOW_97,GUNTER-PH,GUNTER-PD,HMH-APL,MULTI-TEV}. 

The High Altitude Water Cherenkov (HAWC) observatory  is a wide field of view array of 300~water Cherenkov detectors that cover an area of 22,000 $\rm m^2$. HAWC is located at 4100 $\rm m$ above sea level at $19^{\rm o}$ N near the Sierra Negra volcano, in Puebla, Mexico  \cite{HAWC}.
Some effects of LIV are expected to increase with energy and the very long distances due to cumulative processes in the photon propagation (see for instance \cite{CTA-PHz}).
Due to the high altitude and design, HAWC is capable to detect gamma rays with energies between 300 GeV to 100 TeV, which makes it a very useful tool to search LIV signatures in gamma-rays \cite{HAWC-GRB}.  Astrophysical objects as GRB, Pulsars and AGN, could be sources of such high energetic photons. 
Previous works have studied the potential LIV constrains through the possible GRB and Pulsar measurements by HAWC, resulting in strong potential limits to LIV in the photon sector \cite{HAWC-LIV}.
Therefore, in this note we explore the  potential limits that HAWC could provide due to the measurements of very high energy events and the derived consequences of the Lorentz invariant violating photon decay, which can leads to a stringent limit to LIV scale in the photon sector. In Section \ref{LIV}, the highlights of the theory are presented. In Section\ref{sec:limits}, we show the potential limits derived from this method and finally we present the conclusions.

\section{Lorentz Invariance Violation (LIV)}\label{LIV}

A Lorentz violating correction to the dispersion relation can be introduced by a not explicitly Lorentz invariant term at the free particle Lagrangian \cite{GLASHOW}. At the lowest order, the new dispersion relation for a free particle takes the form 
\begin{equation}\label{eq_dis}
	S_a=E_{a}^2-p_a^2 =  m_a^2 \pm \alpha_{a,n} A^{n+2}   \ \  {\rm with} \  \  \alpha_{a,n}=1/(E_{LIV}^{(n)})^n (=\epsilon^{(n)}_{LIV}/ {\rm M}^n)
\end{equation}
for an a-particle with energy and momentum $E$ and $p$. The $\alpha_{a,n}$ factor parametrizes the LIV correction which could be none-universal for all type of particles. $A$ can take the form of $E$, $p$ or a combination of both 
(see for instance Refs. \cite{DIS1,DIS2,DIS3, JACOB, Stecker2009, Stecker2009NJ}).
The effects of LIV are expected to increase with energy and be negligible at the lower standard energies at which 
non LIV-signatures has been found \cite{SME_tables}. To fulfill  such purpose, an energy scale, ${\rm M}$, is introduced. It is common to associate, ${\rm M}$ with $E_{Pl}$, where $E_{Pl}$ is  the Plank energy scale~$\sim$~$10^{19}$~GeV. 
However,  without loss of generality, the LIV correction can be just named $E_{LIV}$, as we do hereafter. 
Several and different techniques have been implemented in the search  of LIV signatures in astroparticle 
physics, and some of them have been used to derive strong constraints to the LIV energy scale,  $E^{(n)}_{LIV}$ (or the LIV-parameter, $\alpha_n$)
see for instance Refs. \cite{ GRB-LIV, HEGRA-LIV, VERITAS-LIV,PULSARS-LIV,HESS-LIV, FERMI-LIV,E1, E2,E3}.

In addition to the very high energy regime,  the very long distances of astrophysical scenarios can lead to a significant LIV effect due to cumulative processes.  That is the case of kinematically forbidden processes that are allowed in LIV scenarios, such as photon decay, photon splitting and spontaneous photon emission (or vacuum cherenkov radiation) \cite{GLASHOW, GUNTER-PH,GUNTER-PD, HMH-APL, MULTI-TEV,VCR-17}. Particularly, photon decay is of special interest for gamma-ray physics, 
since it allows us to use astrophysical sources of VHE photons  as  tests to probe fundamental physics.
Previous works \cite{Proc2,Proc3,HMH-APL} have reported a decay rate for any n-case in the expansion of Eq. (\ref{eq_dis}.) at leading order in the LIV-parameters and for a very high energy photon scenarios, given by

\begin{equation}\label{eq_PD_gamma}
	\begin{aligned}
	\Gamma^{(n)}_{\gamma\rightarrow l^-l^+} =  &  \dfrac{e^2}{4\pi}  \frac{|4m^2 - \alpha_n k^{n+2}| }{4 k\sqrt{1+\alpha_n k^n}} 
	\times  \int_0^{\theta_{max}} \sum_{p_\pm} \frac{p^2\sin\theta d\theta }{| pE' + (p-k\cos\theta)E | }, 
	\end{aligned}
\end{equation}
where

\begin{equation}
{\small
	E=\sqrt{p^2+m^2}, \ \ \ \ E' = \sqrt{k^2 + p^2+m^2-2kp\cos\theta}.
}
\end{equation}
It was also shown that photon decay is allowed after a threshold that depends on the LIV parameter, the photon energy and the mass of the decay products. Above such energy, the decay rate is so efficient that leads to a cut-off in the photon spectrum and no high-energy photons will reach the Earth from cosmological distances. 
There are emission and decay rates on vacuum Cherenkov radiation and photon decay
obtained from different LIV approaches. For instance, expressions from the minimal Standard-Model extension with spontaneous breaking of Lorentz symmetry \cite{SME} and from the introduction of Lorentz violating operators of dimensions four and six can be found in Refs.~\cite{TWO-side,CROSS,Klin_2016}.
In Ref. \cite{HMH-APL} a general expression for the threshold is derived for any  $n$-case in the expansion of Eq. (\ref{eq_dis}.), thus, photon decay is forbidden if
\begin{equation}\label{eq_limit}
    E^{(n)}_{LIV} > E_{\gamma}\left[\frac{E_{\gamma}^2-4 m^2}{4 m^2}\right]^{1/n},
\end{equation}
where $E_{\gamma}$ stands for the high energy photon energy and  $m=m_e$, that is $\gamma \rightarrow e^{+}e^{-}$. 
Hence,  a lower limit for $E_{LIV}$ in the photon sector directly emerges from any observed high energy cosmic photon event.

\section{Limits on LIV}\label{sec:limits}

\begin{figure}[ht!]
	\begin{minipage}{18pc}
	    {\centering
		\includegraphics[width=1.\textwidth]{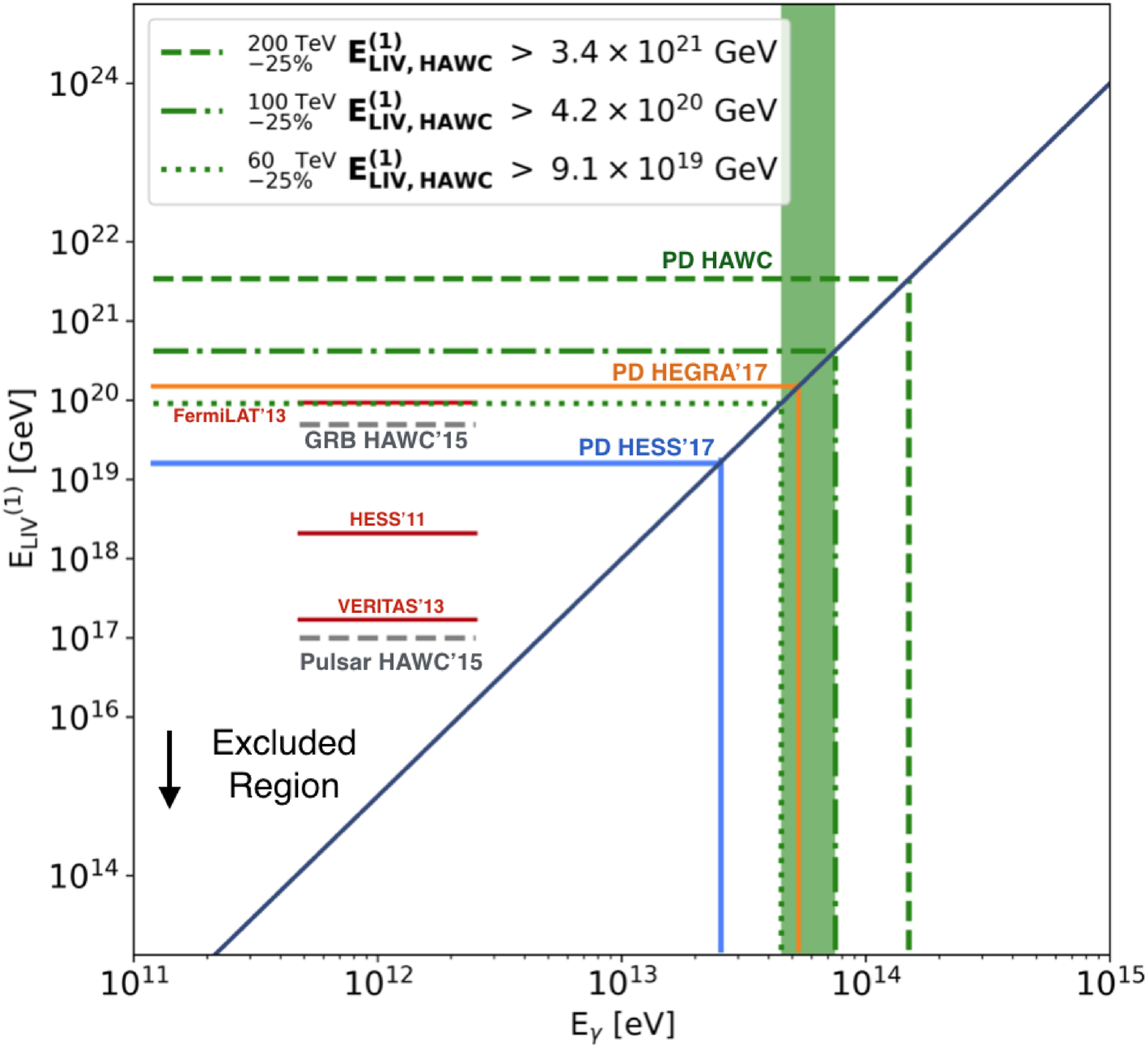}
		}
	\end{minipage}\hspace{0.5pc}%
	\begin{minipage}{18pc}
	    {\centering
		\includegraphics[width=1.\textwidth]{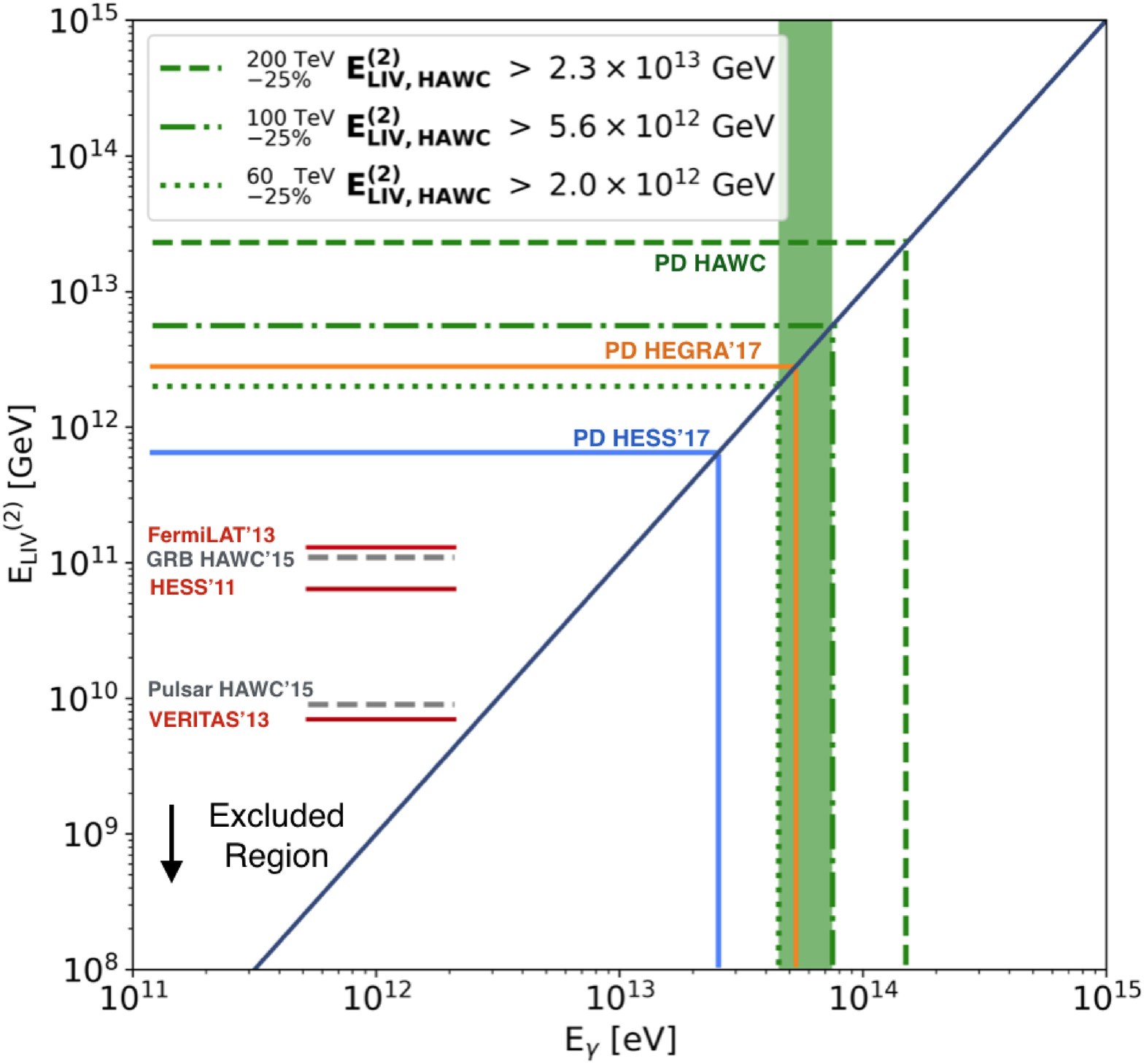}
		}
	\end{minipage} 
	\caption{$E_{LIV}$  excluded region and limits from LIV photon decay into electron positron pairs. In the left $E_{LIV}$  potential limits  from HAWC for n=1 and assuming energy uncertainties of 25$\%$. In the right $E_{LIV}$  potential limits from HAWC for n=2 and assuming energy uncertainties of 25$\%$. For comparison, current limits from other telescopes and different approaches are also shown. Potential limits are displayed in discontinued lines. }\label{fig:1}
\end{figure}

The implementation of the previous generic approach to photon decay on vacuum proves that this process if allowed is very efficient.
The process strongly restricts the possible propagation of the photon to very short distances from source and the outcome of this is a direct and very simple way to bound LIV energy scale that meanly depends in the energy resolution and uncertainties of the detector.
Following this line of thought, assuming a certainty of a photon event and an uncertanty of the $25\%$ in the energy measured by HAWC, potentially compatible with current analysis and energy estimators (see for instance \cite{SAM}), the implementation of  Eq (\ref{eq_limit}) leads to stringent potential limits presented  in Fig.~\ref{fig:1} for $n=$1, 2. Eq (\ref{eq_limit}) is shown in the continuous diagonal (blue) line, it indicates the LIV value, as a function of the VHE photon energy, at which photon decay could be allowed. Thus, the excluded region is given by the measurements of the photon energy, in the horizontal axis, and the value at which it hits the diagonal (blue) line. The green band illustrates the energy uncertainty of the possible 60 TeV event and the lower limit of it is used to limit the region where photon decay is not possible, since photons at that energy  are measured.  For comparison, constrains to the the photon sector by other methods and measurements are shown \cite{HAWC-LIV,VERITAS-LIV,HESS-LIV,FERMI-LIV,HMH-APL}. To 
exemplify the HAWC potential to constrain LIV, three different values are shown: 60, 100 and 200 TeV from bottom to top.  The constrains  
get even more stringent if photons with higher energies are observed.

\section{Conclusions}

The HAWC observatory has been collecting data continuously since 2014 and it was completed in 
March of 2015. Improvements to the energy uncertainties and the understanding of systematics are work in progress. 
However, it can be expected that HAWC measurements can be used as a test to probe fundamental physics such as LIV or set limits to the LIV Energy scale. 

\bigskip

The present work attempt to set the reference for a none conventional but simple method to establish stringent 
limits due to the highest energy photon events detected with HAWC.
As it was discussed,
the High Altitude Water Cherenkov observatory has the potential to set competitive and stringent LIV limits with this and different analysis 
techniques by the accurate observation of very-high energy photons.

\bigskip

\section*{Acknowledgments}

We acknowledge the support from: the US National Science Foundation (NSF); the
US~Department of Energy Office of High-Energy Physics; the Laboratory Directed
Research and De\-ve\-lopment (LDRD) program of Los Alamos National Laboratory;
Consejo Nacional de Ciencia y Tecnolog\'{\i}a (CONACyT), M{\'e}xico (grants
271051, 232656, 260378, 179588, 239762, 254964, 271737, 258865, 243290,
132197), Laboratorio Nacional HAWC de rayos gamma; L'OREAL Fellowship for
Women in Science 2014; Red HAWC, M{\'e}xico; DGAPA-UNAM (grants IG100317,
IN111315, IN111716-3, IA102715, 109916, IA102917); VIEP-BUAP; PIFI 2012, 2013,
PROFOCIE 2014, 2015;the University of Wisconsin Alumni Research Foundation;
the Institute of Geophysics, Planetary Physics, and Signatures at Los Alamos
National Laboratory; Polish Science Centre grant DEC-2014/13/B/ST9/945;
Coordinaci{\'o}n de la Investigaci{\'o}n Cient\'{\i}fica de la Universidad
Michoacana. Thanks to Luciano D\'{\i}az and Eduardo Murrieta for technical support.

{\small{



\bibliography{bibliography}

\providecommand{\newblock}{}
\begin{thebibliography}{10}
\expandafter\ifx\csname url\endcsname\relax
  \def\url#1{{\tt #1}}\fi
\expandafter\ifx\csname urlprefix\endcsname\relax\def\urlprefix{URL }\fi
\providecommand{\eprint}[2][]{\url{#2}}

\bibitem{NAMBU}
Nambu Y 1968 {\em Supplement of the Progress of Theoretical Physics\/} {\bf
  Extra Number} 190--195

\bibitem{Bluhm}
Bluhm R 2014 {Observational Constraints on Local Lorentz Invariance} {\em
  Springer Handbook of Spacetime\/} ed Ashtekar A and Petkov V pp 485--507
  (\textit{Preprint} \eprint{1302.1150})

\bibitem{Pot}
Potting R 2013 {\em J. Phys. Conf. Ser.\/} {\bf 447} 012009

\bibitem{ALFARO}
Alfaro J 2005 {\em Phys. Rev. Lett.\/} {\bf 94} 221302 (\textit{Preprint}
  \eprint{hep-th/0412295})

\bibitem{QG1}
Amelino-Camelia G 2001 {\em Nature\/} {\bf 410} 1065--1067 (\textit{Preprint}
  \eprint{gr-qc/0104086})

\bibitem{QG2}
Ellis J~R, Mavromatos N~E and Nanopoulos D~V 2000 {\em Gen. Rel. Grav.\/} {\bf
  32} 127--144 (\textit{Preprint} \eprint{gr-qc/9904068})

\bibitem{QG3}
Ellis J~R, Mavromatos N~E, Nanopoulos D~V and Volkov G 2000 {\em Gen. Rel.
  Grav.\/} {\bf 32} 1777--1798 (\textit{Preprint} \eprint{gr-qc/9911055})

\bibitem{QG4}
Ellis J~R, Mavromatos N~E and Nanopoulos D~V 2000 {\em Phys. Rev.\/} {\bf D61}
  027503 (\textit{Preprint} \eprint{gr-qc/9906029})

\bibitem{QG5}
Gambini R and Pullin J 1999 {\em Phys. Rev.\/} {\bf D59} 124021
  (\textit{Preprint} \eprint{gr-qc/9809038})

\bibitem{Gian}
Calcagni G 2017 {\em Eur. Phys. J.\/} {\bf C77} 291 (\textit{Preprint}
  \eprint{1603.03046})

\bibitem{SME}
Colladay D and Kostelecky V~A 1998 {\em Phys. Rev.\/} D {\bf 58} 116002

\bibitem{GLASHOW}
Coleman S~R and Glashow S~L 1999 {\em Phys. Rev.\/} {\bf D59} 116008
  (\textit{Preprint} \eprint{hep-ph/9812418})

\bibitem{GLASHOW_97}
Coleman S~R and Glashow S~L 1997 {\em Phys. Lett.\/} {\bf B405} 249--252
  (\textit{Preprint} \eprint{hep-ph/9703240})

\bibitem{GUNTER-PH}
Galaverni M and Sigl G 2008 {\em Phys. Rev. Lett.\/} {\bf 100} 021102
  (\textit{Preprint} \eprint{0708.1737})

\bibitem{GUNTER-PD}
Galaverni M and Sigl G 2008 {\em Phys. Rev.\/} {\bf D78} 063003
  (\textit{Preprint} \eprint{0807.1210})

\bibitem{HMH-APL}
Mart\'inez-Huerta H and P\'erez-Lorenzana A 2017 {\em Phys. Rev.\/} {\bf D95}
  063001 (\textit{Preprint} \eprint{1610.00047})

\bibitem{MULTI-TEV}
Rubtsov G, Satunin P and Sibiryakov S 2017 {\em JCAP\/} {\bf 1705} 049
  (\textit{Preprint} \eprint{1611.10125})

\bibitem{HAWC}
Springer R~W (HAWC) 2016 {\em Nucl. Part. Phys. Proc.\/} {\bf 279-281} 87--94

\bibitem{CTA-PHz}
Fairbairn M, Nilsson A, Ellis J, Hinton J and White R 2014 {\em JCAP\/} {\bf
  1406} 005 (\textit{Preprint} \eprint{1401.8178})

\bibitem{HAWC-GRB}
Alfaro R {\em et~al.\/} (HAWC) 2017 {\em Astrophys. J.\/} {\bf 843} 88
  (\textit{Preprint} \eprint{1705.01551})

\bibitem{HAWC-LIV}
Nellen L (HAWC) 2016 {The potential of the HAWC Observatory to observe
  violations of Lorentz Invariance} {\em {Proceedings, 34th International
  Cosmic Ray Conference (ICRC 2015): The Hague, The Netherlands, July 30-August
  6, 2015}\/} vol ICRC2015 p 850 (\textit{Preprint} \eprint{1508.03930})

\bibitem{DIS1}
Amelino-Camelia G 2001 {\em Nature\/} {\bf 410} 1065--1067 (\textit{Preprint}
  \eprint{gr-qc/0104086})

\bibitem{DIS2}
Ahluwalia D~V 1999 {\em Nature\/} {\bf 398} 199 (\textit{Preprint}
  \eprint{gr-qc/9903074})

\bibitem{DIS3}
Amelino-Camelia G, Ellis J~R, Mavromatos N~E, Nanopoulos D~V and Sarkar S 1998
  {\em Nature\/} {\bf 393} 763--765 (\textit{Preprint}
  \eprint{astro-ph/9712103})

\bibitem{JACOB}
Jacobson T, Liberati S and Mattingly D 2003 {\em Phys. Rev. D\/} {\bf 67}(12)
  124011

\bibitem{Stecker2009}
Scully S~T and Stecker F~W 2009 {\em Astropart. Phys.\/} {\bf 31} 220--225
  (\textit{Preprint} \eprint{0811.2230})

\bibitem{Stecker2009NJ}
Stecker F~W and Scully S~T 2009 {\em New J. Phys.\/} {\bf 11} 085003
  (\textit{Preprint} \eprint{0906.1735})

\bibitem{SME_tables}
Kostelecky V~A and Russell N 2011 {\em Rev. Mod. Phys.\/} {\bf 83} 11--31
  (\textit{Preprint} \eprint{0801.0287})

\bibitem{GRB-LIV}
Xu H and Ma B~Q 2016 {\em Phys. Lett.\/} {\bf B760} 602--604 (\textit{Preprint}
  \eprint{1607.08043})

\bibitem{HEGRA-LIV}
Schreck M 2014 {Obtaining bounds from ultra-high energy cosmic rays in
  isotropic modified Maxwell theory} {\em {Proceedings, 6th Meeting on CPT and
  Lorentz Symmetry (CPT 13): Bloomington, Indiana, USA, June 17-21, 2013}\/} pp
  176--179 (\textit{Preprint} \eprint{1310.5159})

\bibitem{VERITAS-LIV}
Zitzer B (VERITAS) 2013 {Lorentz Invariance Violation Limits from the Crab
  Pulsar using VERITAS} {\em {Proceedings, 33rd International Cosmic Ray
  Conference (ICRC2013): Rio de Janeiro, Brazil, July 2-9, 2013}\/} p 1147
  (\textit{Preprint} \eprint{1307.8382})

\bibitem{PULSARS-LIV}
Otte A~N 2012 {Prospects of performing Lorentz invariance tests with VHE
  emission from pulsars} {\em {Proceedings, 32nd International Cosmic Ray
  Conference (ICRC 2011): Beijing, China, August 11-18, 2011}\/} vol~7 pp
  256--259 (\textit{Preprint} \eprint{1208.2033})

\bibitem{HESS-LIV}
Abramowski A {\em et~al.\/} (H.E.S.S.) 2011 {\em Astropart. Phys.\/} {\bf 34}
  738--747 (\textit{Preprint} \eprint{1101.3650})

\bibitem{FERMI-LIV}
Vasileiou V, Jacholkowska A, Piron F, Bolmont J, Couturier C, Granot J, Stecker
  F~W, Cohen-Tanugi J and Longo F 2013 {\em Phys. Rev.\/} {\bf D87} 122001
  (\textit{Preprint} \eprint{1305.3463})

\bibitem{E1}
Ellis J~R, Mavromatos N~E, Nanopoulos D~V, Sakharov A~S and Sarkisyan E~K~G
  2006 {\em Astropart. Phys.\/} {\bf 25} 402--411 [Erratum: Astropart.
  Phys.29,158(2008)] (\textit{Preprint} \eprint{0712.2781})

\bibitem{E2}
Albert J {\em et~al.\/} (MAGIC, Other Contributors) 2008 {\em Phys. Lett.\/}
  {\bf B668} 253--257 (\textit{Preprint} \eprint{0708.2889})

\bibitem{E3}
Ellis J~R, Mavromatos N~E, Nanopoulos D~V and Sakharov A~S 2003 {\em Astron.
  Astrophys.\/} {\bf 402} 409--424

\bibitem{VCR-17}
Schreck M 2017  (\textit{Preprint} \eprint{1702.03171})

\bibitem{Proc2}
Mart\'inez-Huerta H and P\'erez-Lorenzana A 2016 {\em J. Phys. Conf. Ser.\/}
  {\bf 761} 012035 (\textit{Preprint} \eprint{1609.07185})

\bibitem{Proc3}
Mart\'inez-Huerta H and P\'erez-Lorenzana A 2017 {\em J. Phys. Conf. Ser.\/}
  {\bf 866} 012006 (\textit{Preprint} \eprint{1702.00913})

\bibitem{TWO-side}
Klinkhamer F~R and Schreck M 2008 {\em Phys. Rev.\/} {\bf D78} 085026
  (\textit{Preprint} \eprint{0809.3217})

\bibitem{CROSS}
Rubtsov G, Satunin P and Sibiryakov S 2012 {\em Phys. Rev.\/} {\bf D86} 085012
  (\textit{Preprint} \eprint{1204.5782})

\bibitem{Klin_2016}
Diaz J~S, Klinkhamer F~R and Risse M 2016 {\em Phys. Rev.\/} {\bf D94} 085025
  (\textit{Preprint} \eprint{1607.02099})

\bibitem{SAM}
Marinelli S~S and Goodman J~A (HAWC) 2017 {Measuring High-Energy Spectra of
  Galactic Sources with HAWC.} {\em Proceedings, 35th International Cosmic Ray
  Conference (ICRC2017): Busan, Korea, 10-20 July, 2017.\/} {See
  arXiv:1708.02572 for all HAWC contributions.}

\end{thebibliography}

}}

\end{document}